\begin{filecontents}{mylocalbib.bib}
@misc{Note01,
    note        = "The definition of $H_D$ has been made according to the expression of the volume density free energy associated to DMI given by $\varepsilon_{DMI}=\frac{D}{M_s^2}\left[\left(\mathbf{M}\cdot\mathbf{u}\right)\nabla\cdot\mathbf{M}-\mathbf{M}\cdot\nabla\left(\mathbf{M}\cdot\mathbf{u}\right)\right]$, $\mathbf{u}$ being a unit vector perpendicular to the interface between the ferromagnet and the heavy metal. This definition is coherent with the numerical expressions used by software packages such as MuMax$^3$\cite{Vansteenkiste:14} or GPMagnet\cite{LopezDiaz:12}",
    key         = "note01",
}

\documentclass[10pt,a4paper]{article}
\usepackage[affil-it]{authblk}
\usepackage{amsmath}
\newcommand\sign{\mathrm{sign}}

\usepackage{graphicx}

\begin{document}
\title{Asymmetric driven dynamics of Dzyaloshinskii domain walls in ultrathin ferromagnetic strips with perpendicular magnetic anisotropy.}
\author[1]{L. S\'{a}nchez-Tejerina}
\author[1]{\'{O}. Alejos\thanks{e-mail:oscaral@ee.uva.es}}
\author[2]{E. Mart\'{i}nez}
\author[1]{J. M. Mu\~{n}oz}
\affil[1]{Dpto. Electricidad y Electr\'{o}nica, Facultad de Ciencias, Universidad de Valladolid - 47011 Valladolid, Spain}
\affil[2]{Dpto. F\'{i}sica Aplicada, Facultad de Ciencias, Universidad de Salamanca - 37011 Salamanca, Spain}
\maketitle

\begin{abstract}
The dynamics of domain walls in ultrathin ferromagnetic strips with perpendicular magnetic anisotropy is studied from both numerical and analytical micromagnetics. The influence of the interfacial Dzyaloshinskii-Moriya interaction associated to a bi-layer strip arrangement has been considered, giving rise to the formation of Dzyaloshinskii domain walls. Such walls possess under equilibrium conditions an inner magnetization structure defined by a certain orientation angle that make them to be considered as intermediate configurations between Bloch and N\'{e}el walls. Two different dynamics are considered, a field-driven and a current-driven dynamics, in particular, the one promoted by the spin torque due to the Spin-Hall effect. Results show an inherent asymmetry associated with the rotation of the domain wall magnetization orientation before reaching the stationary regime, characterized by a constant terminal speed. For a certain initial DW magnetization orientation at rest, the rotation determines whether the reorientation of the DW magnetization prior to reach stationary motion is smooth or abrupt. This asymmetry affects the DW motion that can even reverse for a short period of time. Additionally, it is found that the terminal speed in the case of the current-driven dynamics may depend on either the initial DW magnetization orientation at rest or the sign of the longitudinally injected current.
\end{abstract}

\section{Introduction}
Domain walls (DWs) in ferromagnetic materials are boundaries separating regions (domains) that are magnetized differently. The efficient displacement of domain wall (DWs) along thin ferromagnetic strips is a prerequisite condition for the application of the DW-based spintronic devices. \cite{Allwood:05, Parkin:08} DWs in soft ferromagnetic strips, with Permalloy being the most commonly used material, have been extensively analyzed during the last decades, both from theoretical and experimental points of view (see \cite{Boulle:11} for an extended review). Nowadays most of the interest is focused on ultrathin strips sandwiched between a nonmagnetic heavy metal and an insulator, which depict high perpendicular magnetocristalline anisotropy (PMA) and where the current-induced DW motion is anomalously efficient. \cite{Miron:10} Due to the narrow DW widths and lower threshold current densities for motion of DWs, these asymmetric PMA systems are promising platforms for solid-state magnetic devices based on electrically manipulate DWs. 

Apart from their potential for technological applications, and due to the rich physics involved, the analysis of the static properties of DWs in asymmetric stacks with strong PMA, and their field- and current-driven DW dynamics, are both also interesting from a pure fundamental point of view. The high efficiency of the current-induced DW dynamics was initially attributed to a Rashba effective field that stabilizes Bloch DWs against deformation, permitting high-speed motion through conventional non-adiabatic spin-transfer torque (STT). \cite{Miron:10} However, a number of recent findings suggest that STT contributes negligibly to DW dynamics in these ultrathin structures and interfacial phenomena are instead responsible. \cite{Haazen:13,Emori:13,Ryu:13,Torrejon:14} Spin-orbit coupling (SOC) is at the basis of several effects in these systems. Firstly, SOC between the ferromagnetic strip and the insulating oxide overlayer induces the strong PMA in the ferromagnetic strip. Besides, SOC at the interface between the ferromagnetic layer and the heavy metal underlayer may result in different key phenomena explaining the experimental observations.\cite{Haazen:13,Emori:13,Ryu:13,Torrejon:14} Indeed, the spin Hall effect (SHE) in the adjacent heavy metal has emerged as a possible alternative mechanism to the STT. The SHE produces a spin current from charge scattering in the heavy metal, and the resulting spin accumulation at the heavy-metal/ferromagnet interface generates a Slonczeswski-like torque\cite{Liu:12,Liu:12b,Torrejon:15} sufficiently strong to drive the DW motion.\cite{Haazen:13,Emori:13,Ryu:13,Torrejon:14} However, the SHE-induced torque alone cannot directly drive the magnetostatically preferred Bloch DWs in these materials.\cite{Khvalkovskiy:13} The Dzyaloshinskii-Moriya interaction (DMI), arising from the SOC and asymmetric interfaces, determines the magnetization texture of the DWs \cite{Moriya:60,Bode:07,Heide:08,Thiaville:12}. The DMI is a form of magnetic exchange interaction in which adjacent magnetic moments prefer to align orthogonal to each other with a certain handedness, in contrast with the ferromagnetic exchange interaction in which the magnetic moments prefer to align parallel. The DMI provides the missing ingredient to explain the current-induced DW motion stabilizing N\'{e}el DWs with a built-in longitudinal chirality, such that the SHE alone drives them uniformly and with high efficiently. \cite{Emori:13,Thiaville:12,Emori:14,Boulle:13,Martinez:14b,Martinez:14}

Although significant advances have been achieved during the last three years in the analysis of these systems, there exist still some interesting effects which have not studied so far. In particular, in asymmetric PMA heterostructures with moderate DMI interaction, the internal magnetization of the DWs do not depict neither pure Bloch (perfectly aligned along the transverse in-plane axis) nor pure N\'{e}el configurations (perfectly aligned along the longitudinal in-plane axis), but it adopts an intermediate state between them. \cite{Thiaville:12} Here we report an analytical and micromagnetic study which indicates that the internal magnetization moment is indeed degenerated, i.e., under moderate DMI, two DW configurations are energetically possible. While the longitudinal magnetization component is governed by the chirality of DMI, the transverse component is independent of this chirality, so that it can take either a positive or a negative value. This degeneration in the equilibrium state results in asymmetric field-driven and current-driven DW dynamics, which are both analytically and numerically analyzed here in detail.

This work is structured as follows. In section \ref{s:1DM}, the one-dimensional model is used to explore the equilibrium DW configurations at rest with emphasis on describing in details the introduced DW degeneration. The field-driven and the current-driven DW dynamics is evaluated in section \ref{s:Dynamics}, one-dimensional predictions are compared to full micromagnetic simulations. The main conclusions of our study are discussed in section \ref{s:Conclusions}.

\section{One-dimensional model}\label{s:1DM}

The one dimensional model (1DM), as it was originally meant,\cite{Schryer:74, Slonczewski:79} considers the existence of magnetic domains within the medium, separated by domain walls (DWs), transitional areas which establish an interface between two neighbor domains. These DWs can be schematically represented by a certain surface defined implicitly by an analytical function $\psi\left(x,y,z,t\right)=0$.\cite{Consolo:14} In fact, the expression $\psi\left(x,y,z,t\right)=\psi_0$ defines a collection of surfaces that propagates at every point in a direction given by the unit vector $\left.\frac{\nabla\psi}{\left|\nabla\psi\right|}\right|_{\psi=\psi_0}$ at a speed that can be calculated as $\left.\frac{\partial_t\psi}{\left|\nabla\psi\right|}\right|_{\psi=\psi_0}$. 

Since the orientation of the magnetization at every point is expressed by means of a couple of angles which are, in general, dependent on the space coordinates and time, there must exist a couple of extremal values for both angles defining each magnetic domain. In the simplest case, the orientation of the magnetization propagates coherently, and the interface is determined by a constant value of a certain component of the magnetization, which is given by any intermediate value of either one or both orientation angles. The instantaneous position of the interface can be then connected with a certain variable defined all over the interface trajectory.

In the case of large strips, a 1DM can be derived from the Gilbert equation of the magnetization along with the commonly named Walker trial's functions, after the application of variational principles.\cite{Thiaville:02} Walker trial's functions introduce the concepts of DW position, DW magnetization orientation and DW width $\Delta$, the latter obtained in terms of exchange and anisotropy free energies. In the absence of any external torque, one pair of coupled equations can be derived:
\begin{align}
\frac{\dot q}{\Delta}-\alpha\dot\Phi&=\frac{\gamma_0}{2\mu_0 M_s\Delta}\frac{\partial\sigma}{\partial\Phi}\mathrm{,}\\
\alpha\frac{\dot q}{\Delta}+\dot\Phi&=-\frac{\gamma_0}{2\mu_0 M_s}\frac{\partial\sigma}{\partial q}\mathrm{,}
\end{align}
where $\gamma_0=2.21\times 10^5\frac{\mathrm{m}}{\mathrm{A}\cdot\mathrm{s}}$ and $\alpha$ correspond respectively to the gyromagnetic ratio of the free-electron and the Gilbert damping constant, $\mu_0$ is the vacuum permeability, and $M_s$ is the saturation magnetization of the medium. The variable $q$ represents the instantaneous position of the DW along the strip, and $\Phi$ determines in the case of strips with PMA the homogeneous orientation of the in-plane component of the magnetization with respect to the longitudinal direction, as it will be further revised. Finally, $\sigma$ represents the free energy density per unit surface associated to the presence of the DW, which is obtained by integrating the volume free energy density along the longitudinal direction. This free energy density per unit surface originally included exchange, anisotropy and magnetostatic interactions. However, previous equations can be straightforwardly used in order to add Zeeman interactions by considering them as part of such energy density. Some upgrades can also be made to include other external torques,\cite {Hayashi:06,Boulle:14} pinning,\cite{Thomas:07} or thermal effects.\cite{Martinez:12b} Besides, the suitability of the use of the abovementioned trial's functions for strips with interfacial DMI has been adequately stated by other authors,\cite{Thiaville:12} that introduced the concept of Dzyaloshinskii domain walls (DDWs). In this way, DDWs are intermediate cases when the orientation of the in-plane component of the magnetization with respect to the longitudinal direction differs from the extreme situations known as N\'{e}el DWs or Bloch DWs, as it will be further detailed.

In this work, previous equations have been tailored to study the dynamic of deterministic DDWs in perfect strips with PMA driven by both external perpendicular field and SHE, being then written as follows for an up to down transition of the magnetization:\cite{Emori:13}
\begin{align}
\dot\Phi\left(1+\alpha^2\right)&=\gamma_0\left(H_z+\frac{\pi}{2} H_{SH}\cos\Phi\right)+\nonumber\\
&+\alpha\gamma_0\sin\Phi\left(H_K\cos\Phi-H_D\right)\mathrm{,}\label{eq:dotphi}\\
\frac{\dot q}{\Delta}\left(1+\alpha^2\right)&=\alpha\gamma_0\left(H_z+\frac{\pi}{2} H_{SH}\cos\Phi\right)-\nonumber\\
&-\gamma_0\sin\Phi\left(H_K\cos\Phi-H_D\right)\mathrm{.}\label{eq:dotq}
\end{align}
The different $H$-values stand for both the effective fields equivalent to the inner interactions within the ferromagnet, and the external stimuli. In this way, $H_K=M_s\left(N_x-N_y\right)$, where $M_s$ is the saturation magnetization and $N_x$ and $N_y$ are respectively the so-called demagnetizing terms along the longitudinal and transverse axes, whose difference is approximately proportional to the film thickness $t$.\cite{Tarasenko:98} Additionally, $H_D$ is proportional to the Dzyaloshinskii-Moriya parameter $D$ as $H_D=-\frac{\pi D}{2\mu_0 M_s\Delta}$.\cite{Note01}\nocite{Vansteenkiste:14,LopezDiaz:12} Finally, $H_z$ is the applied out-of-plane field and $H_{SH}$ defines the spin-orbit torque (SOT) associated to SHE, depending on the Spin-Hall angle in the heavy metal $\theta_{SH}$\cite{Khvalkovskiy:13} and being proportional to the longitudinally injected current $j_x$ as $H_{SH}=\frac{\hbar\theta_{SH}}{2\mu_0 eM_st}j_x$, with $e$ representing the elementary charge, and $\hbar$ the reduced Planck constant.

\subsection{Equilibrium condition}
Previously to the dynamic characterization, a brief review of the equilibrium condition for DDWs is needed. In the absence of external stimuli the DW magnetization is oriented along a certain angle $\left.\Phi\right|_{eq}=\varphi_0$. The equilibrium condition is then derived from the minimization of the free energy density per unit surface as calculated in \cite{Thiaville:12} 
with regard to the orientation angle. This leads to the equations:
\begin{eqnarray}
\sin\varphi_0\left(H_K\cos\varphi_0-H_D\right)&=0\mathrm{,}\\
H_K\left(\cos^2\varphi_0-\sin^2\varphi_0\right)-H_D\cos\varphi_0&\le 0\mathrm{,}
\end{eqnarray}
that admit a solution in the form:
\begin{equation}
\cos\varphi_0=\sign\left(\frac{H_D}{H_K}\right)\min\left(1,\left|\frac{H_D}{H_K}\right|\right)\mathrm{,}\label{eq:phi0}
\end{equation}
which includes the extreme cases of absence of DMI, that is, $H_D=0$ and $\varphi_0=\pm\frac{\pi}{2}$, as for Bloch DWs, or strong DMI, so that $\left|\frac{H_D}{H_K}\right|\ge 1$, and $\varphi_0=0$ or $\varphi_0=\pi$, as for N\'{e}el DWs. Any other values of the DMI lead to DDWs, so that two orientations of the magnetization within the wall are possible for every single value of the DMI parameter, being these two orientations symmetric with respect to the longitudinal axis. This is schematically depicted in figure \ref{fig.1}, where some axes have been proposed in order to clarify the geometry and the magnitudes here defined. A strip with three micromagnetically calculated domains (up-down-up) is presented, with $m_z$ corresponding to the out-of-plane component of the magnetization, and $m_x$ and $m_y$ being respectively the longitudinal and transverse in-plane components. Details on how micromagnetic calculations have been carried out are further given along the paper. The stacked color maps in each subfigure show the strength of any of these components at the domains and at the DWs. Since the out-of-plane component of the magnetization goes through zero within the DW, the in-plane component take at such point a maximum of magnitude $m_{DW}$ oriented along the $\Phi$ angle. As it has been stated, under equilibrium conditions this angle is fixed to one of the two possible values of $\varphi_0$ which are a solution of eq. \ref{eq:phi0}. In this way, while the longitudinal component of the magnetization within the DW is governed by the chirality due to DMI, the transverse component can take two symmetric values. In the case of figure \ref{fig.1}.a) both DWs have a positive transverse component, while in figure \ref{fig.1}.c) both DWs have a negative transverse component. Different signs of the transverse component for neighbor DWs are also possible, as it is shown in figure \ref{fig.1}.b)

Next section shows that these two degenerated equilibrium states of DDWs for a given value of the DMI parameter present dynamic behaviors under the same external stimulus, such as out-of plane applied fields or SHE associated to longitudinally injected currents, which are intrinsically different. These differences can be also revealed by applying the external stimulus in opposite directions to a certain DDW, as it is further demonstrated.

\begin{figure}
\centering
\includegraphics[height=1.5in]{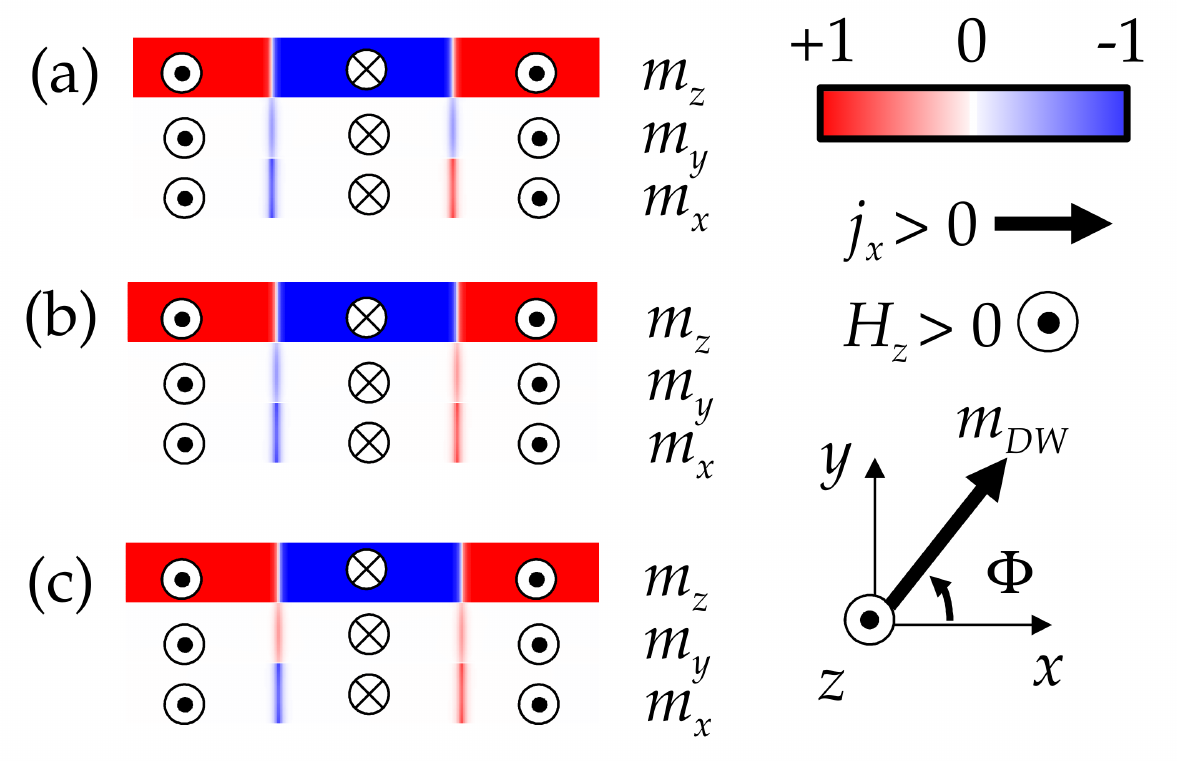}
\caption{\label{fig.1}Definition of the geometry and some of the magnitudes involved in this study, in particular, the applied stimuli in the form of out-of-plane magnetic fields $H_z$ and longitudinal currents $j_x$. Some DDW static configurations are also depicted. In particular, subfigures (a), (b) and (c) show how the chiral character of the DMI forces the longitudinal component of the magnetization $m_x$ within a DW to rotate either clock- or counterclockwise, but not both. However, the transversal component of the magnetization $m_y$ is achiral, so that, this component is free to rotate within the DW. In this way, this rotation can take either the same direction within two consecutive DWs, as in subfigure (b), or different directions, as in subfigures (a) and (c).}
\end{figure}

\section{Dynamics of Dzyaloshinskii DWs}\label{s:Dynamics}
As the above referenced experimental results evidence, the dynamic of a DW admits a stationary motion under the application of certain stimuli, which is characterized by a constant DW speed. The range of the applied stimuli that allows such a stationary behavior depends on the type of torque or interaction. For example, field-driven dynamics is characterized by two different regimes, the abovementioned stationary one for low applied fields, and a lower-velocity regime with a region of negative mobility just above the critical field called Walker field. Nevertheless, current-driven dynamics as this induced by the SHE is characterized by a single and stationary regime. This can be straightforwardly deduced from eq. (\ref{eq:dotphi}), which states that $\dot\Phi$ is a continuous and bounded function of $\Phi$. If the DW starts from an equilibrium condition, the immediate application of either external stimulus gives to $\dot\Phi$ a certain finite value, whose sign defines if the DW magnetization orientation initially rotates clock- or counterclockwise. Two behaviors may occur from this point on. On the one hand, $\dot\Phi$ may vary without a change of sign as $\Phi$ varies in time. This situation occurs when both the upper and lower bounds of $\dot\Phi$ have the same sign, then limiting the rotation speed of the DW magnetization orientation, but preventing the stationary regime from being reached. That is the case of large applied out-of-plane fields, that may mask the other terms in the RHS of eq. (\ref{eq:dotphi}), and Walker breakdown occurs. On the other hand, the rotation of the DW magnetization orientation may reach a certain angle so that $\dot\Phi$ goes through zero. From this instant on, rotation stops and the stationary motion is reached.

According to the previous discussion, it can be stated that the stationary motion is characterized through the conditions $\dot\Phi = 0$, and then $\dot q = const$, the former leading to the relationship:
\begin{equation}
H_z+\frac{\pi}{2} H_{SH}\cos\varphi_s=\alpha\sin\varphi_s\left(H_D-H_K\cos\varphi_s\right)\mathrm{,}\label{eq:phis}
\end{equation}
$\varphi_s$ representing the DW magnetization orientation when the stationary regime is reached. From the comparison between eqs. (\ref{eq:phi0}) and (\ref{eq:phis}), and the characteristics and dependences of $\dot\Phi$ given by eq.(\ref{eq:dotphi}), it can be immediately inferred that the transition from the DW equilibrium state to its stationary motion involves a monotonous rotation of the DW magnetization orientation from $\varphi_0$ to $\varphi_s$.

\subsection{Field-driven dynamics}
As abovementioned, the DDW dynamics driven by the application of an out-of-plane external field admits two different regimes. For the sake of a lighter notation, let us rewrite eq.(\ref{eq:phis}) in the absence of SHE as:
\begin{equation}
h=\sin\varphi_s\left(\delta-\cos\varphi_s\right)\mathrm{,}\label{eq:fdd}
\end{equation}
where $h=\frac{H_z}{\alpha H_K}$ is a normalized value of the applied field, and $\delta=\frac{H_D}{H_K}$ defines the normalized strength of DMI. Since the RHS of previous equation is bounded, there are $h$-values so that the equation has no solution. These $h$-values are then above Walker breakdown, and the frontier can be delimited by considering the global maxima of eq.(\ref{eq:fdd}). At the frontier, the DW magnetization orientation $\varphi_W$ is such so that $\left.\frac{\partial h}{\partial\varphi_s}\right|_{\varphi_s=\varphi_W}=0$. This is a necessary, but not sufficient condition. In any case, this condition leads to the relationship:
\begin{equation}
\cos\varphi_W=\frac{\delta\pm\sqrt{\delta^2+8}}{4}\mathrm{.}\label{eq:cosphiw}
\end{equation}
For strong DMI, characterized by absolute values of $\delta$ greater than one, only one sign in the equation above leads to a valid result, i.e., the minus sign for positive $\delta$, then defining the positive and negative Walker fields. It must be noted that the equilibrium condition for such $\delta$-values establishes pure N\'{e}el walls, that present a perfectly symmetric behavior under the proposed stimuli, as it is further discussed. However, in the case of weak DMI, when DDWs are present, both signs in eq.(\ref{eq:cosphiw}) are possible, one of them corresponds to global extrema while the other correspond to local extrema. As in the case of large $\delta$, global extrema define the limit of the stationary motion under an external applied field, i.e., Walker breakdown.

In order to analyze the importance of local extrema, let us consider a DDW initially in equilibrium with a DW magnetization orientation fulfilling the condition $\cos\varphi_0=\delta$, according to eq.(\ref{eq:phi0}) and the definition of $\delta$. The application of the external field $h$ gives rise to a counterclockwise rotation for positive $h$, that is, applied field along the direction of the magnetization in the up-domain, and a clockwise rotation for negative $h$, as it can be inferred from eq.(\ref{eq:dotphi}). For a certain sign of $h$, and provided $h$ is sufficiently strong, the DW magnetization orientation can reach this local extremum. However, if this $h$-value is fixed, but its sign is changed, no local extrema are reached. This is clearly shown in Figure \ref{fig.2}. Plots represent the DDW dynamics under the influence of an out-of-plane external field. In order to depict realistic values of time, DW position $q$ and DW speed $v$, a value of $\Delta$ of about 6nm has been considered in the graphs. An initial DW magnetization orientation $\varphi_0=60^\mathrm{o}$, corresponding to a $\delta$-value of 0.5, has been taken. Figure \ref{fig.2}.a) is obtained for applied fields of $h=\pm0.23$, slightly higher in absolute terms than the value need to reach a local minimum. A clear asymmetry with the sign of $h$ is shown. While the positive $h$-value promotes a soft transition from the equilibrium condition to the stationary motion, the negative $h$-value gives rise to an abrupt transition of the DW magnetization orientation prior to reach the stationary motion. During this transition, the DDW motion even reverses for a short period of time, as shown in the evolution of the DW position $q$. Similar results can be obtained if the applied field is increased close to the Walker breakdown (Figure \ref{fig.2}.b). However, the shown asymmetry do not affect to the terminal speed which is given by $\gamma_0H_z\Delta$.

\begin{figure*}
\includegraphics[height=1.5in]{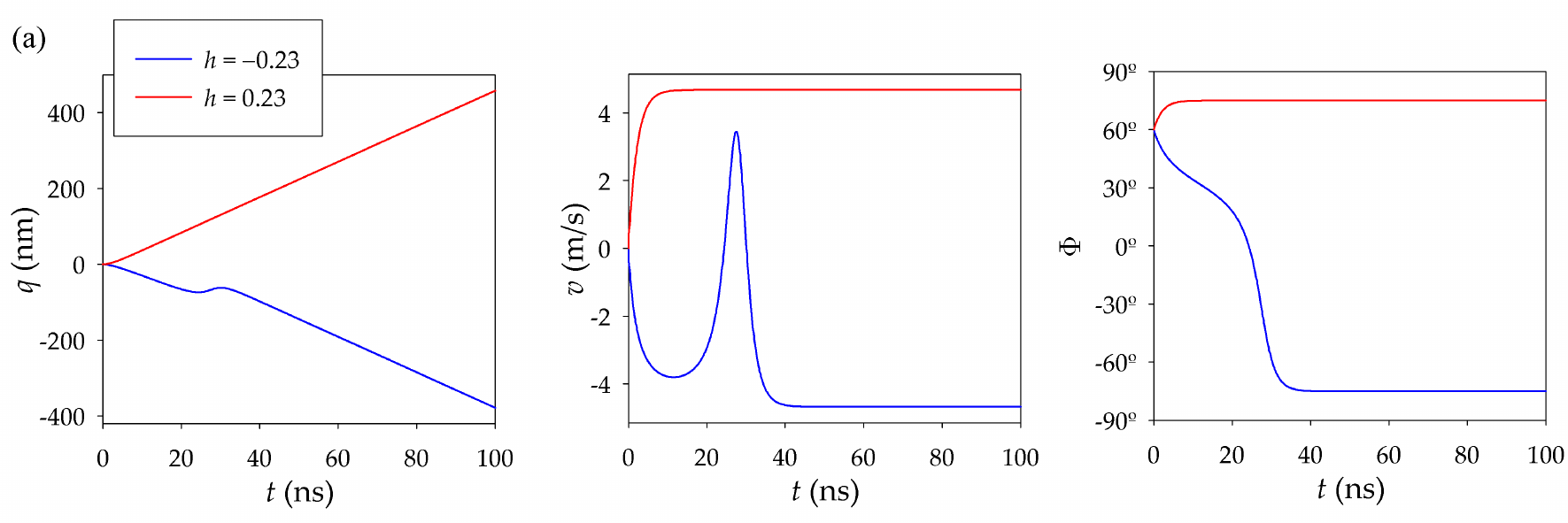}
\includegraphics[height=1.5in]{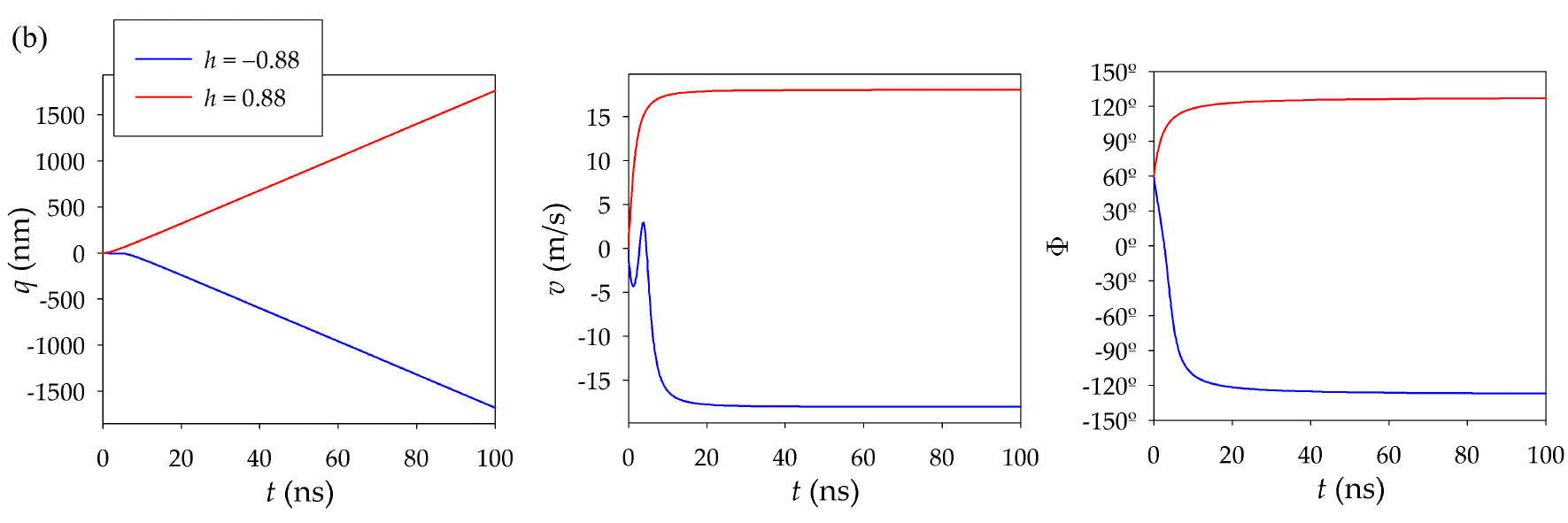}
\caption{DDW dynamics under the influence of an out-of-plane external field. In these graphs, $h$ represents a normalized value as it has been defined along the text. Positive $h$-values stand for fields applied along the direction of the magnetization in the up-domain, while negative values stand for fields applied in the opposite direction, i.e., the magnetization in the down-domain. $q$ represents the instantaneous DDW position, and $v$ is the instantaneous speed. $\Phi$ corresponds to the DW magnetization orientation. Plots correspond to the numerical calculation of eqs.(\ref{eq:dotq}) and (\ref{eq:dotphi}). Figure a) is obtained for applied fields of $h=\pm0.23$, and figure b) is obtained for applied fields of $h=\pm0.88$. The sign of the applied field may promote completely different DDW dynamics.}
\label{fig.2}
\end{figure*}

Micromagnetic simulations have been carried out with the help of the GPMagnet software package\cite{LopezDiaz:12} in order to support these analytical results. Along these simulations, a ferromagnetic strip of 1nm thickness and 160nm width with PMA grown on a heavy metal layer has been considered. Values of the material parameters that are usually found in the literature have been taken: saturation magnetization of $M_s=7\cdot 10^5\frac{\mathrm{A}}{\mathrm{m}}$, exchange constant of $A=10^{-11}\frac{\mathrm{J}}{\mathrm{m}}$, uniaxial anisotropy constant $k_u=4.8\cdot 10^5\frac{\mathrm{J}}{\mathrm{m}^3}$. A DMI parameter $D$ of the order of $10^{-5}\frac{\mathrm{J}}{\mathrm{m}^2}$ has been used.

As an example, Figure \ref{fig.3} presents the micromagnetic results obtained in the case of a sample with $\delta=-0.5$. In this case, equilibrium conditions determine that one of the two possible orientations at rest of the DW magnetization is $\varphi_0=120^\mathrm{o}$, as it can be checked in the graph for $h=0$. The figure shows the dependence of the DW magnetization orientation at stationary motion as a function of the external field $h$. Dots are calculated by means of micromagnetic simulations, while the continuous plot corresponds to the 1DM analytical treatment. In this example, the stationary motion is smoothly reached from the equilibrium orientation for negative fields, up to the corresponding negative Walker field. However, an abrupt transition occurs for positive $h$-values, before the Walker field is reached. The inset shows the normalized in-plane components of the magnetization at the center of the DDW. $m_x$ is the component along the longitudinal axis, while $m_y$ is the transverse component. Starting from equilibrium conditions (filled dot), negative fields give rise to a smooth clockwise rotation of the DW magnetization orientation for all applied fields up to the Walker field. On the contrary, positive fields leads to a counterclockwise rotation that pass through an abrupt reorientation of the DW magnetization prior to reach Walker breakdown.

\begin{figure}
\includegraphics[width=2.5in]{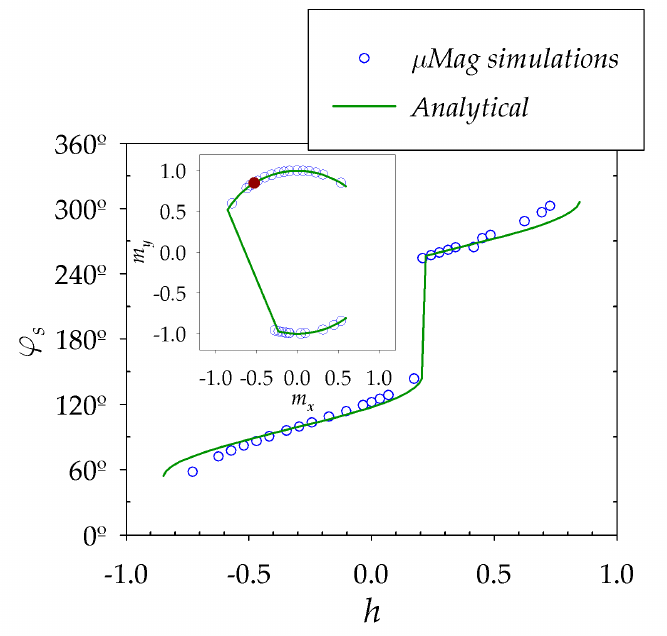}
\caption{Comparison between micromagnetic simulations and the predictions of the 1DM analytical model. Different signs of the applied fields lead to different behaviors of the a DDW.}
\label{fig.3}
\end{figure}

In order to complete the 1DM analytical study, Figure \ref{fig.4} shows the dependence of the DW magnetization orientation at stationary motion on the normalized out-of-plane field $h$ with $\delta$, i.e., $\frac{H_D}{H_K}$, as a parameter. On the one hand, large $\frac{H_D}{H_K}$ ratios promote N\'{e}el walls, their chirality depending on the sign of the DMI parameter $D$, whose behavior is completely symmetric with regard to the sign of $h$. As it is well-known, the larger the ratio is, the farther the two endpoints of the corresponding curve are from each other, meaning that Walker breakdown requires larger applied fields. On the other hand, no DMI allows the existence of Bloch walls. In this case, the dependence of the DW magnetization orientation at stationary motion is depicted for a Bloch DW with $\varphi_0=90^\mathrm{o}$. The behavior is also symmetric with regard to the sign of $h$. However, DDWs present a clear asymmetry under the application of an external out-of plane field. This behavior can be considered as hysteretic in the following sense. Under equilibrium conditions, a DDW may take two possible orientations, symmetric with respect to the longitudinal axis. The application of an out-of-plane field promotes the rotation of the initial DW magnetization orientation before a state of stationary motion is reached. Depending on the sign and the strength of the applied field, the DW magnetization orientation may undergo an abrupt reorientation prior to the stationary state. If this reorientation occurs, the DDW do not recover its original orientation at rest, but the symmetric one with respect to the longitudinal axis, after field removal. The existence or absence of these DW magnetization reorientation noticeably defines DDW dynamics, as the dynamic plots in Figure \ref{fig.2} showed.

\begin{figure}
\includegraphics[width=2.5in]{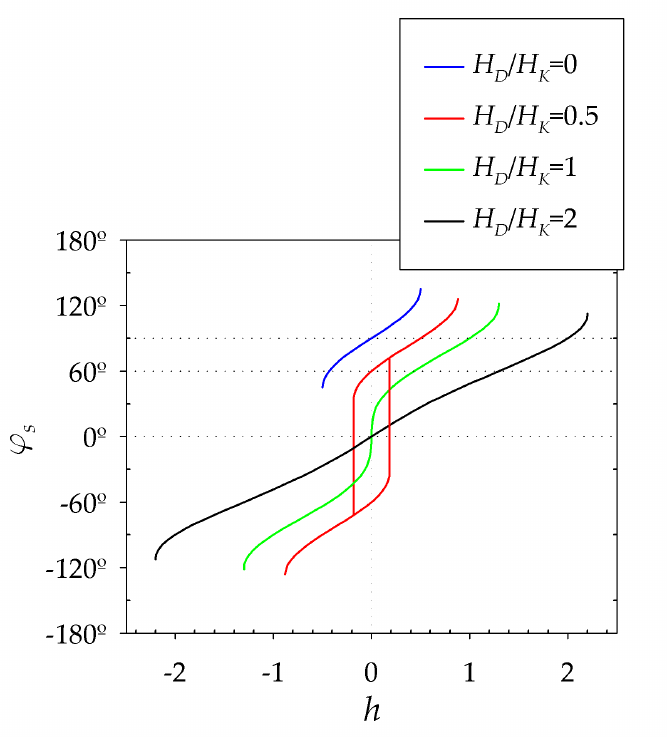}
\caption{Dependence of the DW magnetization orientation at stationary motion on the applied out-of-plane field with $\frac{H_D}{H_K}$ as a parameter. While both Bloch and N\'{e}el walls present symmetric behavior, DDWs present a clear asymmetry with the applied field governed by the strength of the DMI.}
\label{fig.4}
\end{figure}

\subsection{Current-driven dynamics}
As it has been already stated, field-driven dynamics and current-driven dynamics governed by SHE present a noticeable difference, that is, the absence for the latter of Walker breakdown. This can be straightforwardly proved from eq.(\ref{eq:phis}). Analogous to the precedent study, a lighter version of this equation can be proposed in the following way:
\begin{equation}
h=\tan\varphi_s\left(\delta-\cos\varphi_s\right)\mathrm{,}\label{eq:cdd}
\end{equation}
where now $h=\frac{\frac{\pi}{2}H_{SH}}{\alpha H_K}$. Since the RHS of previous equation is not bounded, a stationary motion is always possible under the SHE torque. However, the function defined in eq. (\ref{eq:cdd}) may possess local extrema for absolute $\delta$-values ranging from 0 to 1, that is, for DDWs. This local extrema are reached when the DW magnetization orientation fulfills the condition $\cos\varphi_L=\sqrt[3]{\delta}$. As in the precedent case, when the DW magnetization orientation reaches either of the two local extrema, an abrupt reorientation occurs prior to the establishment of the stationary motion. 

As a first example, Figure \ref{fig.5} shows the DDW dynamics under the SHE for a certain current flowing longitudinally in both one and the opposite direction calculated by means of the 1DM analytical description. In order to give a much more clear idea about orders of magnitude, one set of concrete values has been chosen, which are current density $j_x=\pm 4\cdot 10^{10}\frac{\mathrm{A}}{\mathrm{m}^2}$, Spin-Hall angle $\theta_{SH}=0.11$, with a damping constant $\alpha=0.013$ and a DW width $\Delta$ of about 6nm, values that can be also found in the referenced literature. In a similar fashion to the field-driven dynamics, a smooth rotation of the DW magnetization is promoted if the current flows in one direction, while an abrupt transition occurs when the current flows in the opposite direction.
\begin{figure*}
\includegraphics[height=1.5in]{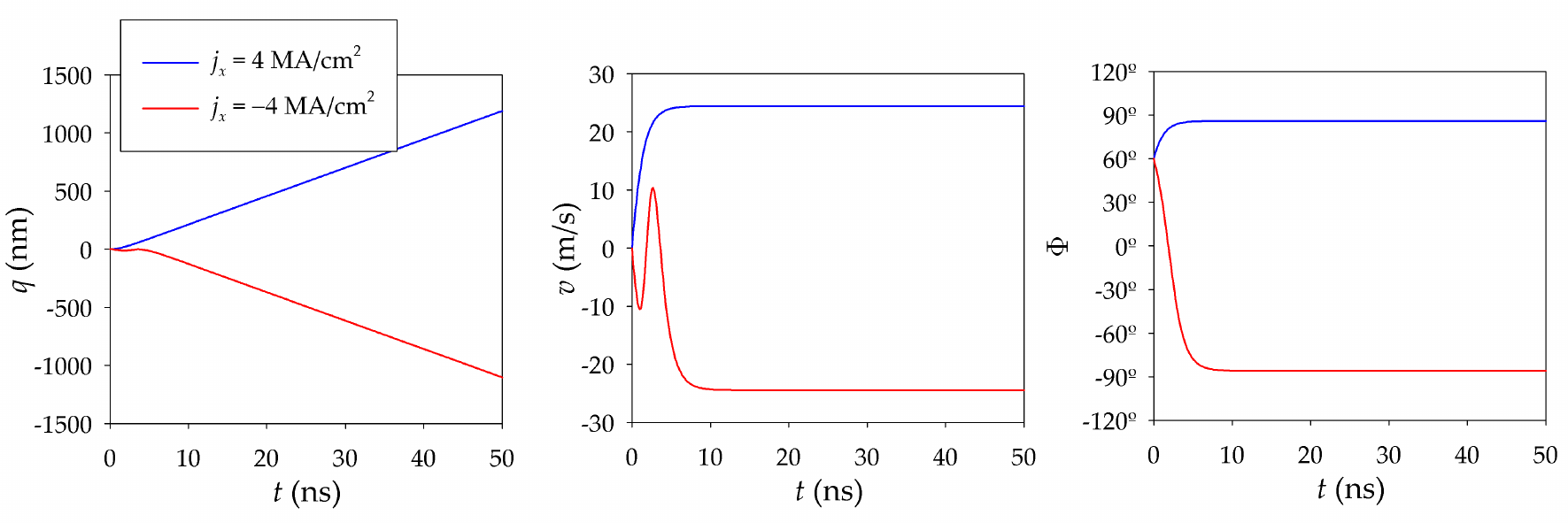}
\caption{Current-driven DDW dynamics under the influence of SHE for positive and negative longitudinal currents as calculated analytically with the help of the 1DM. As in the case of the field-driven dynamics, asymmetry with the sign of the stimulus is also noticeable.}
\label{fig.5}
\end{figure*}

Differently from the field-driven dynamics, the 1DM also predicts the asymmetry of the terminal speed at stationary motion, which undergoes a sharp transition for currents that lead the DW magnetization orientation to values close to the local extrema. Figure \ref{fig.6} shows the dependence of this terminal speed on the $h$-value defined after eq.(\ref{eq:cdd}). Dots correspond to numerical simulations carried out with GPMagnet using the abovementioned parameters, while the continuous plot has been analytically calculated from the 1DM. A rather good agreement between numerical and analytical calculations is obtained. Both show the sharp transition predicted by the 1DM, in this case, for an $h$-value around -20. It can be noticed that the terminal speed at the peak of this sharp variation, calculated from the 1DM as $\frac{\gamma_0\Delta\sqrt[3]{\delta}}{\alpha}\frac{\pi}{2}H_{SH}$, may surpass the terminal speed for infinite current given by $\gamma_0H_D\Delta$.

\begin{figure}
\includegraphics[width=2.5in]{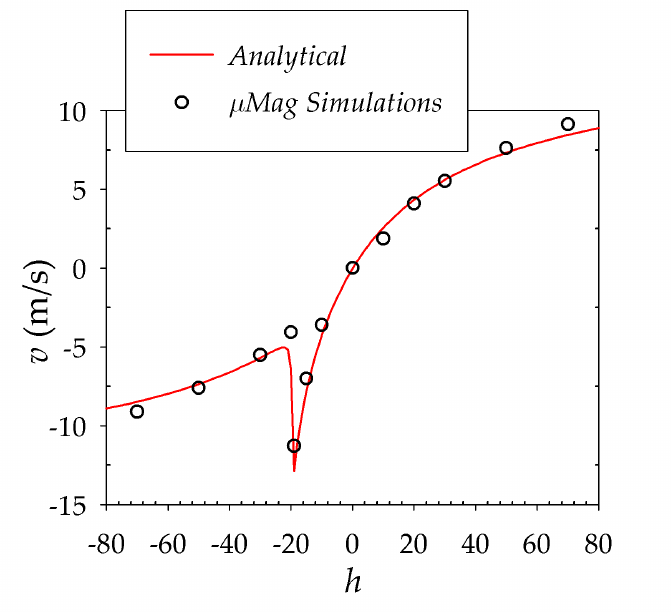}
\caption{Dependence of the DDW terminal speed at stationary motion on the applied longitudinal current due to SHE. The $h$-value in the graph  is defined as proportional to the current density (see text). The graph compares micromagnetic simulations and the results obtained from the 1DM model. A noticeable asymmetry is found if the stimulus is reversed, leading to a sharp transition of the terminal speed in a certain range of applied currents.}
\label{fig.6}
\end{figure}

As a summary, the analytically calculated dependence of the DW magnetization orientation at stationary motion on the applied current with $\frac{H_D}{H_K}$ as a parameter is depicted in Figure \ref{fig.7}. The case of Bloch DWs is not considered, since SHE do not promote their motion in the system under study. For $\frac{H_D}{H_K}$ ratios greater or equal to one, the plots show that N\'{e}el walls behave symmetrically under the influence of the torque due to SHE. However, DDWs have a hysteretic-like behavior, rather analogous to the one found for the field-driven dynamics. It must be noticed that, according to this hysteretic-like behavior, the terminal speed depends on the initial DW magnetization orientation under equilibrium conditions, i.e., for the same applied longitudinal current, the dynamics depends on whether the initial DW magnetization orientation at rest is pointing in one of the two possible directions, symmetric with respect to the longitudinal axis, that this orientation can take.

\begin{figure}
\includegraphics[width=2.5in]{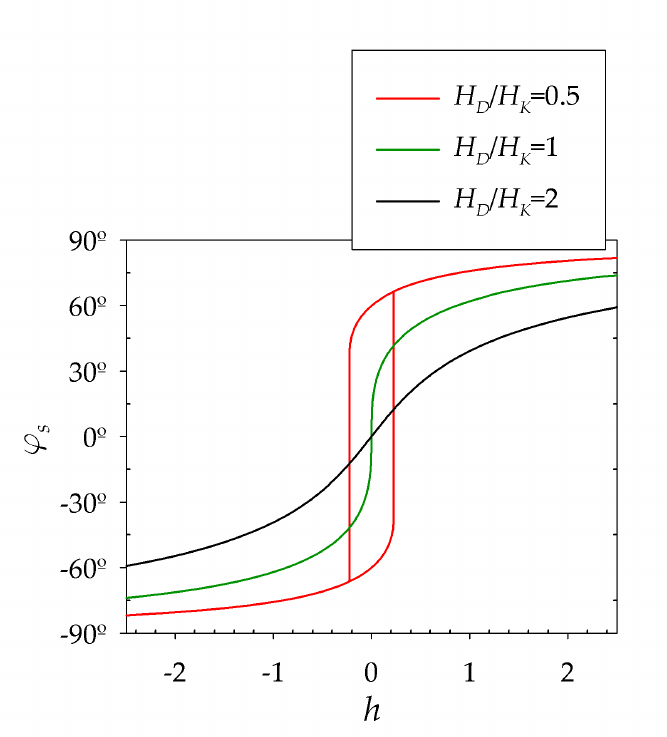}
\caption{Dependence of the DW magnetization orientation at stationary motion on current through the normalized $h$-value (see text) with $\frac{H_D}{H_K}$ as a parameter. It is shown that N\'{e}el walls present symmetric behavior, while DDWs present a hysteretic-like behavior. As a consequence, a DDW may reach different terminal speeds depending on either the sign of the current or its initial orientation at rest.}
\label{fig.7}
\end{figure}

\section{Conclusions}\label{s:Conclusions}
The dynamics of Dzyaloshinskii DWs in ferromagnetic strips with PMA has been analytically studied with the help of the 1DM, and the results compared with micromagnetic simulations. Two different stimuli have been considered: out-of-plane applied magnetic fields and torques induced by longitudinal currents due to SHE. The application of such stimuli may lead to a stationary regime characterized by a constant terminal DW speed. However, the behavior of the DDW dynamics has been found to be asymmetric with respect to either the sign of the applied stimulus, or the initial DW magnetization orientation. This asymmetry arises from the fact that, depending on this sign, the DW magnetization orientation may rotate clock- or counterclockwise. The stationary regime is then reached when this rotation stops. However, the transition from the DW magnetization orientation at rest to the orientation at stationary motion can be either smooth or abrupt depending on the rotation. The initial DW magnetization orientation under equilibrium conditions then determines which sign of the stimulus promotes its abrupt reorientation prior to reach the stationary regime. This process can be characterized in the presented results by a reverse of the DDW motion for a short period of time.

Additionally, some differences between field-driven and current-driven dynamics have been reported. Some of them have been already highlighted by other authors, such as the absence of Walker breakdown in the case of current-driven dynamics governed by SHE. Besides, the asymmetry here reported do not affect to the DW terminal speed in the case of field-driven dynamics. However, the terminal speed for current-driven dynamics is found to be asymmetric in the range of lower injected currents, then depending on either the initial DW magnetization orientation or the sign of the applied stimulus. 

Finally, since for higher injected currents the abrupt reorientation of the DW magnetization occurs, the degenerated condition of the initial DW magnetization orientation also determines different traveled distances for a train of current-driven DWs in a strip, which may be also of experimental relevance.

\section{Acknowledgment}
This work was supported by project WALL, FP7-PEOPLE-2013-ITN n. 608031 from European Commission, project MAT2014-52477-C5-4-P from Spanish Government and project SA282U14 from Junta de Castilla y Le\'{o}n.

\bibliography{mmag,mylocalbib}
\bibliographystyle{unsrt}
\end{document}